\documentclass[12pt]{article}

\usepackage[sort&compress,numbers,super]{natbib}
\usepackage{times}
\usepackage{graphicx}
\usepackage{amsmath}
\usepackage[running]{lineno}
\usepackage{color}
\usepackage[table]{xcolor} 
\definecolor{hgrau}{rgb}{0.7,0.7,1.2} 
\usepackage{array}
\newcolumntype{M}[1]{>{\centering\arraybackslash}m{#1}}

\newcounter{lastnote}

\title{Field-effect control of Graphene-Fullerene thermoelectric nanodevices}

\author
{
	Pascal Gehring,$^{1\ast}$ Achim Harzheim,$^{1}$ Jean Spi\`{e}ce,$^{2}$ Yuewen Sheng,$^{1}$\\ Gregory Rogers,$^{1}$ Charalambos Evangeli,$^{2}$ Aadarsh Mishra,$^{1}$ Benjamin\\ Robinson,$^{2,3}$ Kyriakos Porfyrakis,$^{1}$ Jamie H. Warner,$^{1}$ Oleg Kolosov,$^{2}$\\  G. Andrew. D. Briggs,$^{1}$ Jan A. Mol$^{1}$\\
\\
\normalsize{$^{1}$Department of Materials, University of Oxford, 16 Parks Road, Oxford OX1 3PH, UK}\\
\normalsize{$^{2}$Physics Department, Lancaster University, Lancaster LA1 4YB, UK}\\
\normalsize{$^{3}$Materials Science Institute, Lancaster University, Lancaster, LA1 4YW, UK}\\
\\
\normalsize{$^\ast$To whom correspondence should be addressed; E-mail:  pascal.gehring@materials.ox.ac.uk.}
}

\date{}

\begin{document}

\baselineskip24pt

\maketitle

\section*{Abstract}
\textbf{Although it was demonstrated that discrete molecular levels determine the sign and magnitude of the thermoelectric effect in single-molecule junctions, full electrostatic control of these levels has not been achieved to date. Here, we show that graphene nanogaps combined with gold micro-heaters serve as a testbed for studying single-molecule thermoelectricity. Reduced screening of the gate electric field compared to conventional metal electrodes allows controlling the position of the dominant transport orbital by hundreds of meV. We find that the power factor of graphene-fullerene junctions can be tuned over several orders of magnitude to a value close to the theoretical limit of an isolated Breit-Wigner resonance. Furthermore our data suggests that the power factor of isolated level is only given by the tunnel coupling to the leads and temperature. These results open up new avenues for exploring thermoelectricity and charge transport in individual molecules, and highlight the importance of level-alignment and coupling to the electrodes for optimum energy-conversion in organic thermoelectric materials.}
\section*{Introduction}
The thermopower or Seebeck coefficient $S$ of a material or nanoscale device is defined as $S= -\Delta V/\Delta T$, where $\Delta V$ is the voltage difference generated between the two ends of the junction when a temperature difference $\Delta T$ is established between them. In addition to the goal of maximising $S$, there is a great demand for materials with a high power factor $S^2G$, which is a measure for the amount of energy that can be generated from a temperature difference, and high thermoelectric efficiency, which is expressed in terms of a dimensionless figure of merit $ZT = S^2GT/\kappa$, where $T$ is the average temperature, $G$ is the electrical conductance and $\kappa$ is the sum of the electronic and phononic contribution to the thermal conductance. In conventional thermoelectric materials $S$, $G$ and $\kappa$ are typically mutually contra-indicated, such that high $S$ is accompanied by low $G$ and high $G$ by high $\kappa$\cite{Heremans2013}. In some nanostructured materials these properties can be decoupled\cite{Rama2001}. Therefore, the thermoelectric properties of nanostructures like carbon nanotubes\cite{Small2003}, quantum dot devices\cite{Staring1993,Svensson2012, Svensson2013}, and single-molecule junctions \cite{Kim2014,Evangeli2013,Reddy1568,Baheti2008,Malen2009,Yee2011,Widawsky2012,Garcia2016} have been studied extensively. In the past few years it has been demonstrated both experimentally and theoretically that, at the molecular scale, $S$ can be controlled by the chemical composition\cite{Baheti2008}, the position of intra-molecular energy levels relative to the work function of metallic electrodes\cite{Yee2011}, by systematically increasing the single-molecule lengths within a family of molecules\cite{Reddy1568,Malen2009}, and by tuning the interaction between two neighbouring molecules\cite{Evangeli2013}. Despite these advances, single-molecule experiments have only yielded values of $S$ ranging from 1 to 50 $\mu$V~K$^{-1}$\cite{Kim2014,RinconGarciaCSR2016}. The key challenge in achieving high Seebeck coefficients in molecular junctions lies in controlling the energetic position and ``steepness'' of the transport resonances.

We use graphene-based lateral single-molecule devices -- where a molecule sits in the gap between two graphene leads -- to study the gate-dependent thermoelectric properties of C$_{60}$ molecules. The two-dimensional nature of graphene electrodes leads to a reduced screening of the gate electric field compared to bulky metal electrodes\cite{Lortscher2013}, enabling us to shift the orbital energy levels of the molecule with respect to the electrochemical potential of the graphene leads using a back-gate. We exploit this field-effect control to map the thermo-voltage across entire molecular transport resonances. 
\section*{Experimental part}
\begin{figure}[!ht]
	\centering
	\includegraphics[width=1\textwidth]{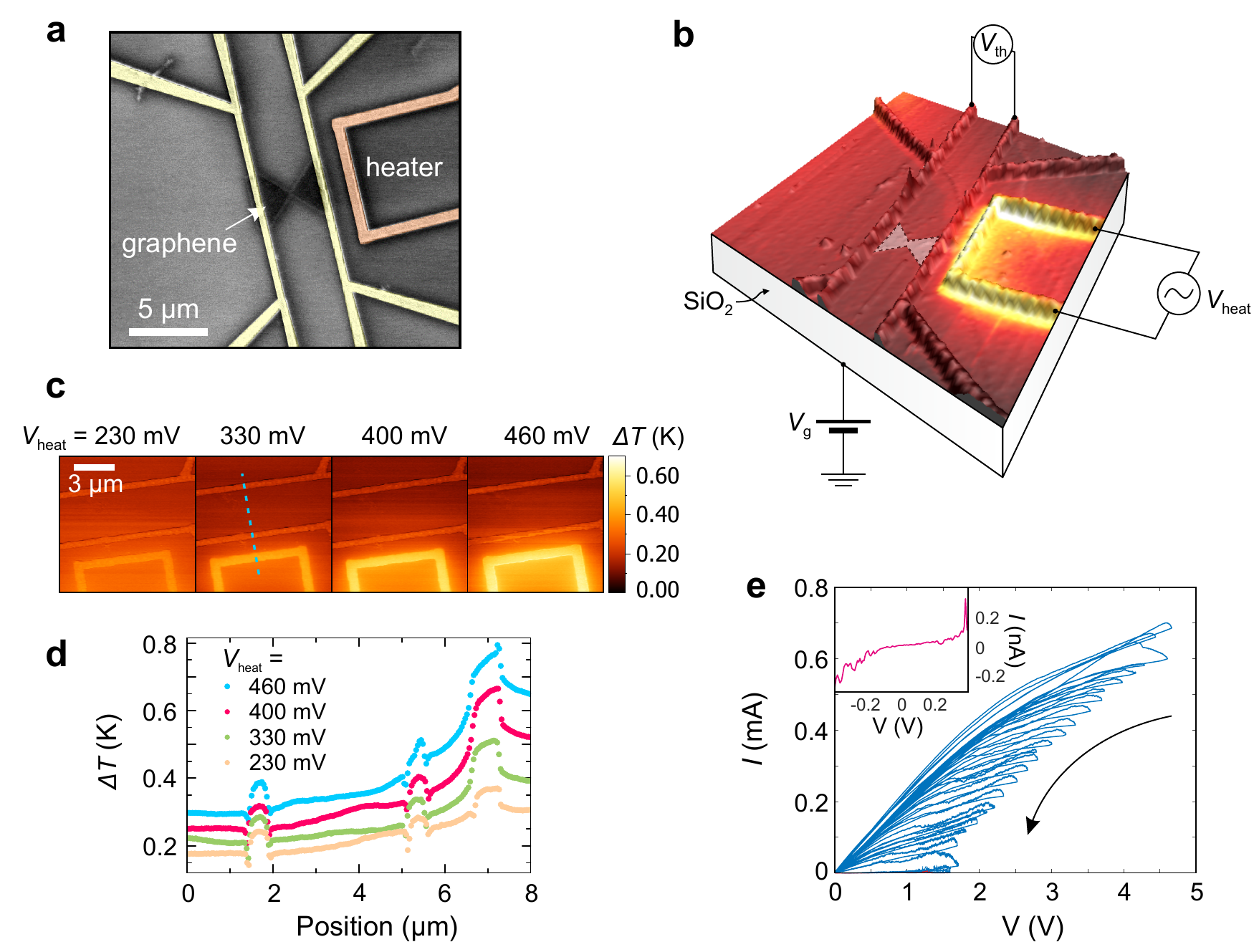}
	\caption{\textbf{Device geometry and Scanning Thermal Microscopy. a}, False-colour scanning electron microscopy image of the device. \textbf{b}, Atomic force microscopy height profile overlaid with scanning thermal microscopy signal and sketch of the device geometry for a typical thermo-voltage measurement. \textbf{c}, Scanning thermal microscopy images recorded at different constant voltages $V_{\text{heat}}$ applied to the micro-heater. \textbf{d}, Line profiles along the device extracted from the maps shown in \textbf{c} (see blue dotted line). \textbf{e}, $I–V_{\text{sd}}$ traces recorded during feedback-controlled electroburning. Inset: $I – V_{\text{sd}}$ trace after completed electroburning.}
	\label{fig:Figure1}
\end{figure}

Our devices consist of CVD graphene etched into bow-tie shape on-top of gold contacts (see Methods for fabrication details). Each gold lead has four contacts for precise 4-terminal resistance measurements, which allows us to measure the temperature difference across the graphene junction (see Figure S1). A gold micro-heater is fabricated 1 $\mu$m away from the junction (see Figure \ref{fig:Figure1}a). 
 By passing a current through the micro-heater we create a temperature gradient across the junction\cite{Small2003,Zuev2009,Devender2016}. We quantify this temperature gradient by  cross-checking several methods to eliminate potential systematic errors. These are: (i) measuring the resistance of the left and right gold contacts; (ii) using COMSOL finite-element simulations; and (iii) using Scanning Thermal Microscopy (SThM) measurements. Using method (i) we measure a temperature difference between the hot (closer to the micro-heater) and cold (further from the micro-heater) contact as a function of heater power $\Delta T/P_{\text{heater}} = 58 \pm 11$~K~W$^{-1}$ at $T_0 = 77$~K (see Chapter 1 and 8 Supporting Information for details of the calibration method and an estimation of the total uncertainty, respectively). This is in close agreement with the finite-element simulations (method (ii)) which predict $\Delta T/P_{\text{heater}} = 50$~K~W$^{-1}$ and a constant temperature gradient $\nabla T/P_{\text{heater}} = 14$~K~$\mu$m$^{-1}$~W$^{-1}$ across the length of the graphene junction (see Figure S5). Figure \ref{fig:Figure1}b shows a temperature map overlaid onto a height profile that were simultaneously recorded using a SThM (method (iii)). 
 From the temperature maps recorded for different heater powers in Figure \ref{fig:Figure1}c and d we extract a power-dependent temperature gradient $\nabla T/P_{\text{heater}} = 18$~K~$\mu$m$^{-1}$~W$^{-1}$ and a temperature difference $\Delta T/P_{\text{heater}} = 63 \pm 10 $~K~W$^{-1}$ between the two gold contacts under ambient conditions and $\Delta T/P_{\text{heater}} = 71 \pm 11 $~K~W$^{-1}$ for 77~K and vacuum (see Chapter 3 Supporting Information). For all the analysis presented below we will use the value extracted using method (i).

We use feedback-controlled electroburning\cite{Prins2011,Lau2014,Puczkarski2015,Gehring2016} (see Figure \ref{fig:Figure1}e) to first form graphene nano-gaps suitable for characterisation of single molecules\cite{Mol2015} in which we subsequently couple C$_{60}$ molecules functionalised with pyrene anchor groups (see Figure \ref{fig:Figure2}a). We have chosen this molecule since it is stable in air, has previously been successfully coupled to graphene electrodes\cite{Lau2016}, and because its thermoelectric properties have been studied using various other techniques, including STM based break junctions\cite{Yee2011,Evangeli2013} and electromigrated gold break junctions\cite{Kim2014}. After electroburning we characterise the graphene gaps by measuring the current $I_{\text{sd}}$ as a function of gate and bias voltage (stability diagram) at $T_0=77$ K in vacuum. Empty devices, where there are no carbon islands or ribbons bridging the gap, are characterised by non-linear $I_{\text{sd}} - V_{\text{sd}}$ curves and little or no gate modulation. After this first characterisation step we warm up the device and deposit C$_{60}$ molecules by immersing the sample in a 10~$\mu$M chloroform solution containing the C$_{60}$ bisadducts for 1~min followed by blow drying with nitrogen gas. We then measured the devices again at low temperature to look for signatures of molecules. In total we fabricated 1080 two-terminal devices on which we performed feedback controlled electroburning. Due to limitations of our setup we were then only able to study 100 devices at low temperatures of which 16 devices showed signatures of molecule deposition: 1) a clear change from ``empty'' to Coulomb blockade after molecule deposition; 2) vibrational fingerprints in the excited state spectrum measured in the sequential tunneling regime.
\section*{Results and discussion}
We often observe multiple overlapping, non-closing Coulomb diamonds which indicate the formation of molecular junctions where more than one molecule contribute to the electrical transport. In the following we discuss the data for 3 selected devices where the Coulomb diamonds close in the accessible back-gate region with addition energies $> 400$~meV. We focus on these devices as their transport is most likely dominated by a single molecule. Moreover, the large addition energies enables us to study well isolated energy levels that are expected to show the largest Seebeck coefficient. Chapter 5 of the Supporting Information includes the data of all measured devices.
\begin{figure}[!ht]
	\centering
	\includegraphics[width=0.5\textwidth]{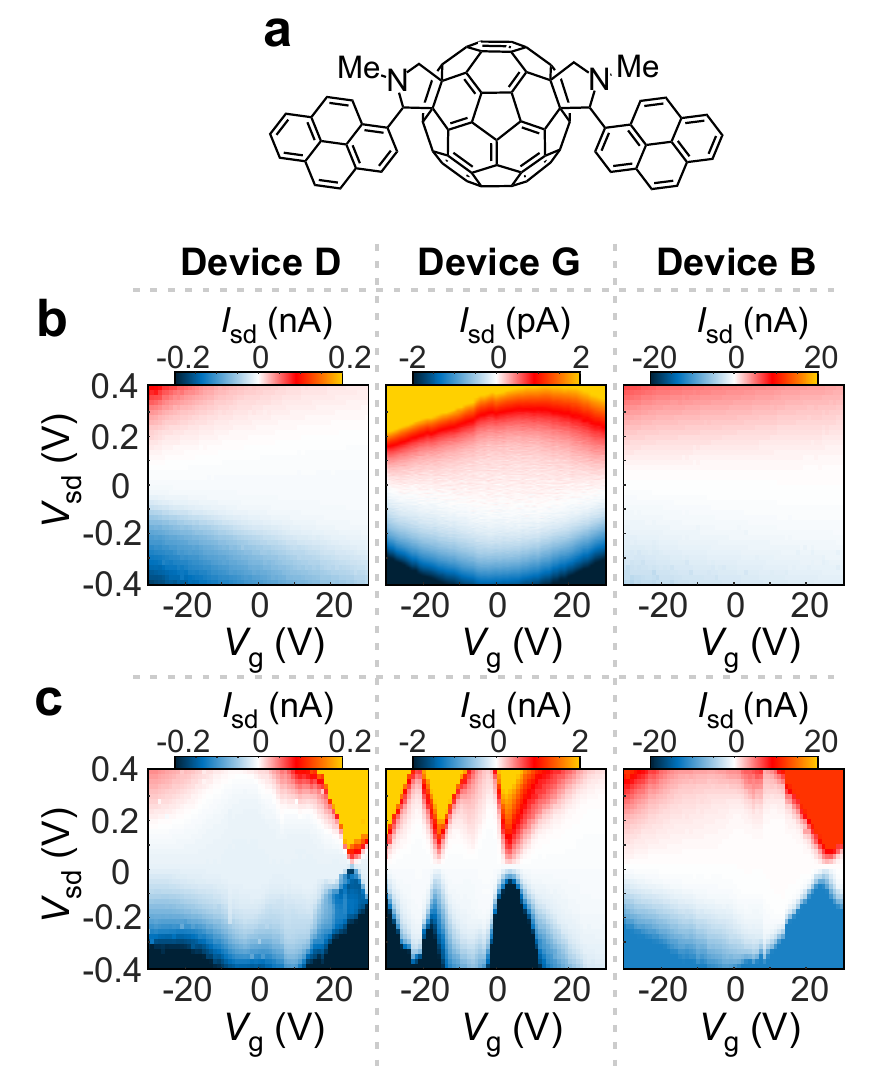}
	\caption{Electrical characterisation of Graphene-Fullerene Single-Molecule thermoelectric nanodevices. (a) Chemical structure of the C$_{60}$ bisadducts functionalised with pyrene anchor groups. (b) Current map as a function of back gate and bias voltage before and (c) after molecule deposition recorded at $T_0 = 77$~K for devices D, G and B, respectively.}
	\label{fig:Figure2}
\end{figure}

Figure \ref{fig:Figure2}b shows the stability diagrams of devices B, D and G measured at 77~K before and after molecule deposition. In Figure \ref{fig:Figure2}b, before molecule deposition, the source-drain current shows only weak gate dependence, but in Figure \ref{fig:Figure2}c regions of Coulomb blockade can be observed after deposition. We attribute the sequential electron tunneling after molecule deposition to the formation of a molecular junction.\cite{Prins2011,Lau2016,Mol2015}. To further investigate the single electron transport, we  studied several devices at low temperatures ($T < 5$~K). In a previous study we observed exited state lines in the sequential tunneling regime that correspond to $H_g(1)$ and $A_g(1)$ Raman active vibrational modes of C$_{60}$ as well as centre-of-mass motion of the C$_{60}$ molecule with respect to the graphene electrodes.\cite{Lau2016}. In total, 7 of 16 devices showed similar evidence for vibrational excited states (see Table S2 Supporting Information). 4 out of 16 devices changed permanently to a non-conducting state after cool down to $<5$~K and no low-temperature data could be recorded. The visibility of vibrational excited states strongly depends on temperature, the tunnel coupling to the leads\cite{Schinabeck2016} and the Franck-Condon factors\cite{Park2000,Burzuri2014} which can vary drastically between different molecular junctions and the charge-transition investigated.\cite{Leon2008,Lau2016} Moreover, density of states fluctuations in the graphene leads can lead to features inside the sequential tunneling regime, which do not run parallel to the edges of the Coulomb diamonds, that can obscure any vibrational fingerprint.\cite{Gehring2017}

\begin{figure}[!ht]
	\centering
	\includegraphics[width=1\textwidth]{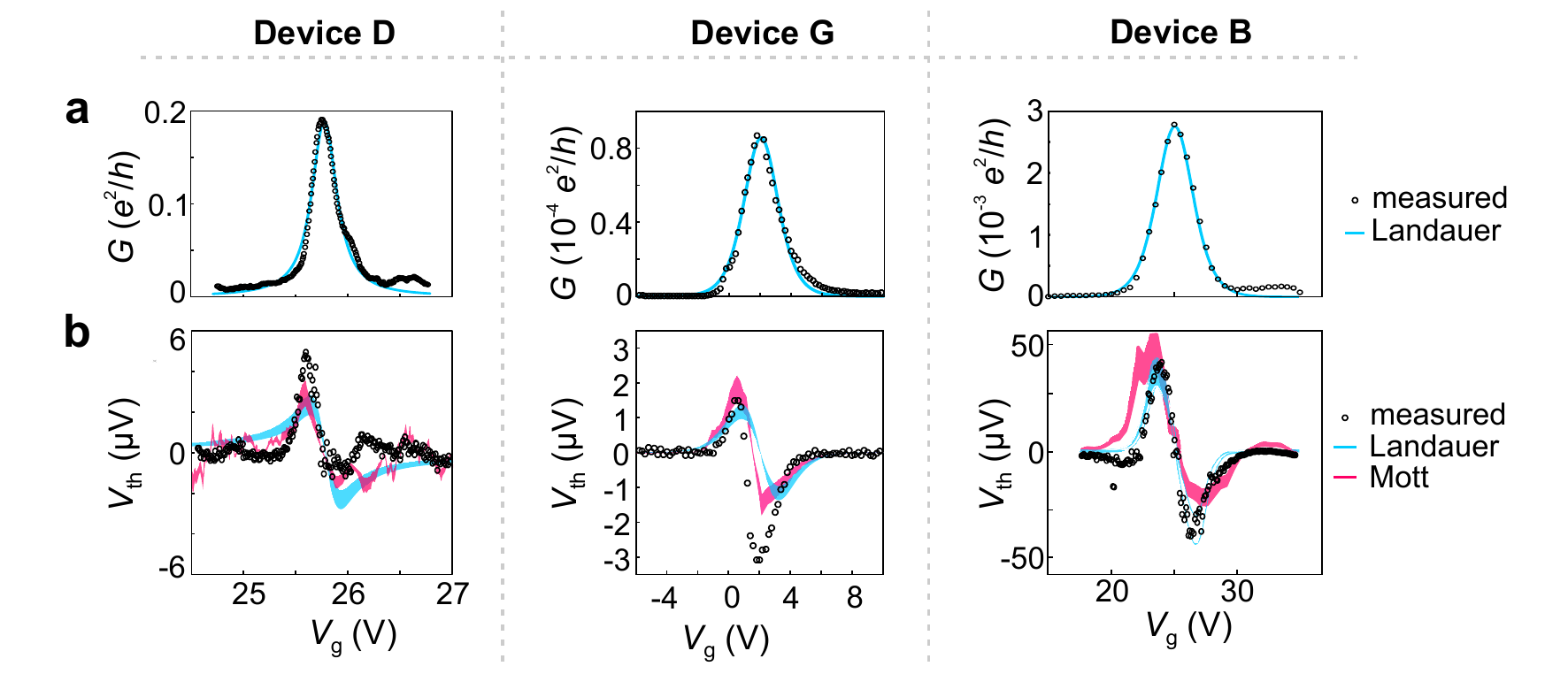}
	\caption{\textbf{Thermoelectric measurements.} (a) AC Zero-bias conductance and (b), Thermovoltage (measured with a temperature bias of $\Delta T= 45 \pm 9$~mK, $\Delta T= 100 \pm 20$~mK and $\Delta T= 180 \pm 36$~mK, respectively) as a function of back gate voltage measured at $T_0 = 3.2$~K (Device D) and $T_0 = 77$~K (Device G and B). The blue and pink curves show theoretical calculations using a Landauer-type approach and the Mott formula, respectively.}
	\label{fig:Figure3}
\end{figure}

Next, we measure the gate dependent thermoelectric properties of the C$_{60}$-graphene junctions. We apply an AC-voltage with modulation frequency $f$ to the micro-heater and measure the thermo-voltage $V_{\text{th}}$ drop on the device at a frequency $2f$ for different back gate voltages $V_\text{g}$ (see Figure \ref{fig:Figure1}b).\cite{Zuev2009} We focus on the high-conductance gate region around the Coulomb peaks (see gate traces in Figure \ref{fig:Figure3}a) since the thermo-voltage signal inside the Coulomb blocked region is smaller than the noise level of our measurement setup. Figure \ref{fig:Figure3}b shows the measured gate-depended thermo-voltage signal for Device D, G and B, recorded at $\Delta T= 45 \pm 9$~mK, $\Delta T= 100 \pm 20$~mK and $\Delta T= 180 \pm 36$~mK, respectively. An increase of $V_\text{th}$ followed by a sign change, further decrease and subsequent increase towards zero can be observed. Similar results have been observed for 7 other devices (see Chapter 5 Supporting Information). Using the applied temperature bias $\Delta T$ we find maximum Seebeck coefficients $S_{\text{max}}$ ranging from 1.5 to 460~$\mu$V~K$^{-1}$ (see Table \ref{tab:Tab1}). On average, these values are more than one order of magnitude larger than the Seebeck coefficients found in STM break junction experiments of C$_{60}$ contacted with different metal electrodes\cite{Yee2011,Evangeli2013}. In the following we use a simple model for an isolated Breit-Wigner resonance to explain these results.


In the linear temperature and bias regime the conductance $G$ can be expressed in terms of the moments $L_i$ of the transmission coefficient $P(E)$ as\cite{Finch2009}
\begin{equation}
G(V_g,T_0) = \frac{2e^2}{h}L_0
\label{eq:conductance}
\end{equation}
with
\begin{equation}
L_i = \int_{-\infty}^{\infty} \left(E - E_F\right)^i ~P(E) ~dE,
\label{eq:moments}
\end{equation}
where we use the non-normalised probability distribution\cite{Lambert2015}
\begin{equation}
P(E) = -\mathcal{T}(E) \frac{\partial f(E)}{\partial E}.
\label{eq:probdist}
\end{equation}
For a single, well isolated molecular level we can assume a Breit-Wigner resonance to describe the transmission probability $\mathcal{T}(E)$:
	\begin{equation}
	\mathcal{T}(E) = \frac{\Gamma_\text{L}\Gamma_\text{R}}{ \left( \Gamma_\text{L}/2 + \Gamma_\text{R}/2 \right)^2 + \left[\left(e\alpha V_\text{g} - E_0\right) - E\right]^2 },
	\label{eq:BW}
	\end{equation}
where $E_0$ is the energy of the transport resonance, $\Gamma_\text{L}$, $\Gamma_\text{R}$ are the tunnel couplings to the leads, and the lever arm $\alpha = \frac{\text{d}E}{\text{d}V_{\text{g}}}$ is determined by the capacitive coupling of the molecule to the gate, source and drain electrodes\cite{Hanson2007}. The derivative of the Fermi-Dirac distribution is
\begin{equation}
\frac{\partial f(E)}{\partial E} = \frac{1}{4kT_0} \cosh^{-2}\left(\frac{E}{2kT_0} \right).
\label{eq:diffFermi}
\end{equation}


In the limit where $\Gamma = \Gamma_\text{L} + \Gamma_\text{R}\gg k_{\text{B}}T_0$ Equation (1) reduces to $G = \frac{2e^2}{h}\mathcal{T}(E)$, and the tunnel coupling to the two leads can be inferred from the height and width of the Coulomb peak. In the opposite limit where $\Gamma \ll k_{\text{B}}T_0$ the maximum conductance $G_{\text{max}}$ is proportional to $\frac{\Gamma_\text{L} \Gamma_\text{R}}{\Gamma_\text{L} + \Gamma_\text{R}}$ while the width of the Coulomb peak is proportional to $k_{\text{B}}T_0$.

\begin{table}[!h]
	
	\caption{Measurement results for each C$_{60}$ device.}
	\centering
	\label{tab:Tab1}
	\begin{tabular}{|M{1.5cm}|M{1.5cm}|M{1.5cm}|M{1.5cm}|M{1.5cm}|M{1.5cm}|M{1.5cm}|M{1.5cm}|M{1.5cm}|}
		
		\hline
		
		\textbf{Device name} & \textbf{$\Gamma$ ($\mu$eV)} & \textbf{$\chi$} & \textbf{$\alpha$ (meV/V)} & \textbf{$E_0$ (meV)} & \textbf{$S_\text{max}$ ($\mu$V/K)} & \textbf{$G_\text{max}$ ($e^2/h$)} & \textbf{$(S^2 G)_\text{max}$ ($k_\text{B}^2/h)$} & $T_0$ (K)  \\
		
		\hline
		B & 88  & --  & 9 & 221 & 220 & 0.003 & 0.01 & 77  \\
		\hline
		C & $1.3 \times 10^3$ & -- & 10 & 188 & 140 & 0.08 & 0.08 & 77   \\
		\hline
		D & $2.7 \times 10^3$ & 15 & 13 & 335 & 27 & 0.2 & 0.14 & 3.2  \\
		\hline
		E & 16 & -- & 6 & 53 & 238 & 0.006 & 0.02 & 11 \\
		\hline
		F & $1.7 \times 10^2$ & -- & 11 & 84 & 460 & 0.01 & 0.11 & 77  \\
		\hline
		G & 2 & -- & 9 & 12 & 30 & $10^{-4}$ & 0.04 & 77   \\
		\hline
		Q & $2.4 \times 10^4$ & $1.2 \times 10^4$ & 62 & 564 & 1.5 & $2 \times 10^{-4}$ & $7 \times 10^{-4}$ & 77  \\
		\hline
	\end{tabular}
\end{table}

When a temperature bias $\Delta T$ is applied to a junction, the Fermi-Dirac distribution of the hot contact broadens compared to that of the cold contact. This gives rise to a thermal current $I_\text{th}$, which leads to a thermo-voltage $V_{\text{th}}$ when measured under open circuit conditions $I(\Delta T, V_{\text{th}})=0$. The ratio of the thermo-voltage and the temperature drop is the Seebeck coefficient $S = - V_{\text{th}} / \Delta T$. Similar to the conductance, the Seebeck coefficient is given by a Landauer-type expression using Equation \ref{eq:moments}, \ref{eq:probdist} and \ref{eq:diffFermi}:
\begin{equation}
S(V_g,T_0) = - \frac{1}{eT_0}\frac{L_1}{L_0}.
\label{eq:ColinSeebeck}
\end{equation}
If $\mathcal{T}(E)$ varies only slowly with $E$ on the scale of $k_BT_0$, i.e. $\Gamma\gg k_BT_0$, then $S$ takes the well-known form of the Mott approximation\cite{Lunde2005}
\begin{linenomath*}
	\begin{equation}
	S = -\frac{\pi^2 k_{\text{B}}^2 T_0}{3 e \alpha} \frac{1}{G} \frac{\text{d}G}{\text{d} V_{\text{g}}},
	\label{eq:Mott}
	\end{equation}
\end{linenomath*}

In Figure \ref{fig:Figure3}b we compare our experimental results to the calculated thermo-voltages using the Mott approximation (equation \ref{eq:Mott}) and the Landauer-type approach (equation \ref{eq:ColinSeebeck}), respectively, where the width of the curve indicates the error in estimating the temperature drop on the junction (see full error analysis in Chapter 8 Supporting Information). For both calculations the thermo-voltage was corrected by a damping factor due to the input impedance of the voltage amplifier (see Chapter 8.4 Supporting Information)\cite{Svensson2012}. To compare the measured thermo-voltage to that obtained from Equation \ref{eq:ColinSeebeck} and \ref{eq:Mott} we assume that the temperature difference $\Delta T$ between the hot and the cold side of the molecule is equal to the temperature difference measured between the two gold contacts. Since cooling lengths of up to $7 \mu$m have been reported for graphene\cite{Song2011}, the assumption that hot electrons injected from the gold contacts into the graphene leads do not thermalise before they reach the junction area approximately 1.7 $\mu$m away from the gold contacts is justified. By assuming that no temperature drops on the graphene leads we only estimate a lower bound of $S$. In addition, we neglect the effect of thermo-voltages created in the strongly p-doped graphene leads whose Seebeck coefficient is on the order of 10~$\mu V/K$\cite{Zuev2009}. However, this would result in a small, constant offset of the thermo-voltage in the applied gate voltage regime far away from the Dirac point of our graphene,\cite{Gehring2016} which we do not observe in our experiments.

For the calculation of the Seebeck coefficient using the Landauer-type approach (equation \ref{eq:ColinSeebeck}) we estimate $\mathcal{T}(E)$ by equation \ref{eq:BW} and extract the tunnel coupling by fitting the gate-dependent conductance traces to Equation \ref{eq:conductance} if $\Gamma\gg k_{\text{B}}T_0$. For those devices where $\Gamma\ll k_{\text{B}}T_0$, we estimate $\mathcal{T}(E)$ by fitting the conductance data with a thermally broadened conductance peak $G = G_{\text{max}} \cosh^{-2}\left[  (\alpha V_\text{g} - E_0)/(2k_\text{B}T_0) \right]$ with $G_{\text{max}} = e^2/(\hbar 4 k_\text{B}T) \Gamma_{\text{L}}\Gamma_{\text{R}}/(\Gamma_{\text{L}}+\Gamma_{\text{R}})$,\cite{Beenakker1991} where we fix $T_0 = 77$~K, and find a lower bound for $\Gamma$ by taking $\Gamma_{\text{L}}=\Gamma_{\text{R}}$ such that $\Gamma_{\text{lower}} = 4 k_\text{B}T_0\frac{2h}{\pi e^2}G_{\text{max}}$. Despite the fact that we can not uniquely determine $\Gamma$ in this regime, there is still good agreement between the measured and calculated thermo-voltage curves. This is due to the relative insensitivity of $S$ on the lifetime of the transport resonance when $\Gamma\ll k_{\text{B}}T_0$ (see Figure S22).

\begin{figure}[!ht]
	\centering
	\includegraphics[width=0.5\textwidth]{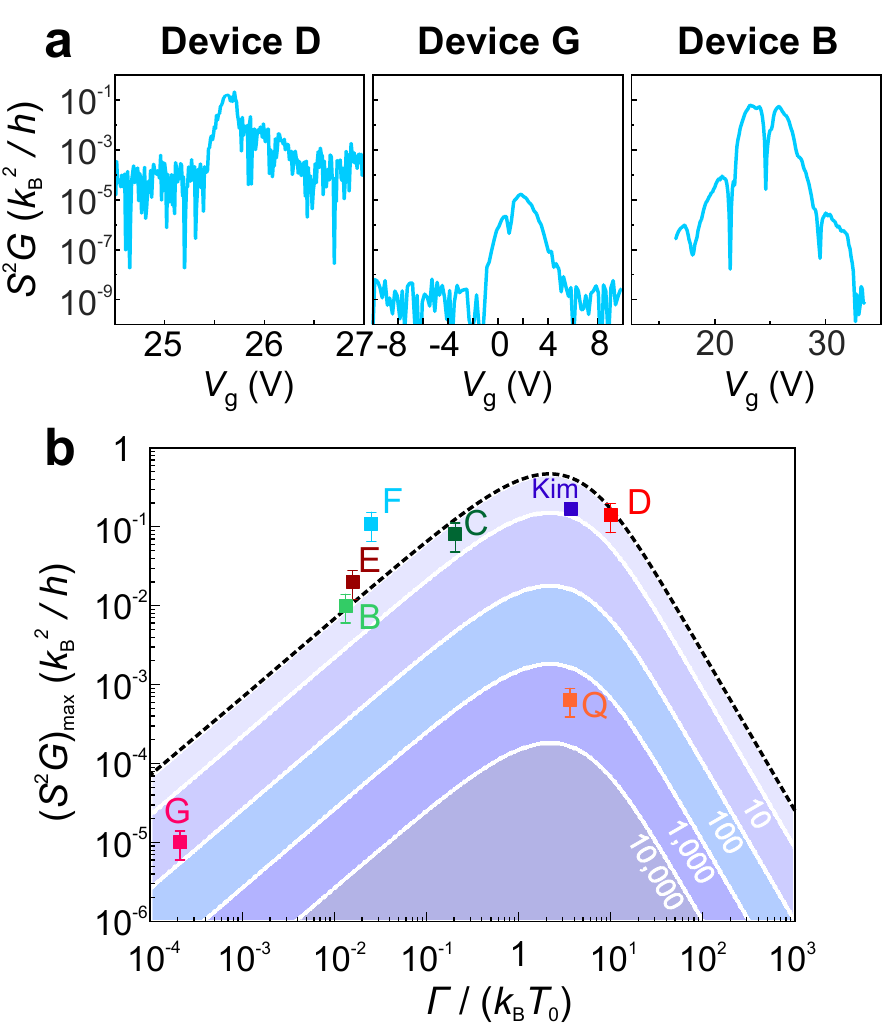}
	\caption{\textbf{Maximum power factor.} (a) Power factor as a function of back gate voltage measured at $T_0 = 3.2$~K (Device D) and $T_0 = 77$~K (Device G and B). (b) Maximum power factor as a function of tunnel coupling $\Gamma$ for the devices investigated in this study and by Kim \textit{et al.}\cite{Kim2014}. The dashed black line ($\chi = 1$) and the white lines show theoretical curves calculated using Equation \ref{eq:PF} for different ratios between the fast and slow tunnel rates $\chi$. The error bars of the data points are estimated by neglecting the error in $G$ and using the relative error in determining $\Delta T$ (20 \%) to estimate the error for $S$. The total error of the power factor $PF = S^2G$ is propagated: $\Delta PF = \sqrt{\left( \frac{\partial PF}{\partial S} \Delta S \right)^2} = 2 S G \Delta S = 2S^2 G \frac{\Delta S}{S}$.}
	\label{fig:Figure4}
\end{figure}
Finally, we use our experimental results to calculate the power factor $S^2G$ for Devices D, G and B (see Figure \ref{fig:Figure4}a and Chapter 5 Supporting Information for other devices). Significantly, we find that $S^2G$ can be tuned by several orders of magnitude by electrical gating to maximum values of $0.01 -  0.14 \times $~$k_\text{B}^2/h$ (see Table \ref{tab:Tab1}). These values are one to two orders of magnitude larger than values found in C$_{60}$ junctions without sufficient electric field control\cite{Yee2011,Evangeli2013,Garcia2016} and comparable to the value found for C$_{60}$ measured using gold break junctions with an electrical back gate\cite{Kim2014}.


To evaluate the thermoelectric performance of different devices, we plot the maximum power factor $(S^2 G)_\text{max}$ on a log-log scale as a function of the temperature-normalized tunnel rate $\Gamma/k_{\text{B}}T_0$. We compare these values to the theoretical maximum calculated using
\begin{linenomath*}
	\begin{equation}
	S^2 G = \frac{2}{h T_0^2}\frac{L_1^2}{L_0},
	\label{eq:PF}
	\end{equation}
\end{linenomath*}
and Equations \ref{eq:conductance} - \ref{eq:ColinSeebeck}. In addition to the theoretical maximum power factor for devices with symmetric tunnel coupling (black dashed line), we plot the theoretical values for $(S^2 G)_\text{max}$ for different ratios between the fast and slow tunnel rates $\chi = \Gamma_{\text{fast}}/\Gamma_{\text{slow}}$ (solid white lines), where $\Gamma_{\text{fast}} = \text{max}(\Gamma_{\text{L}}, \Gamma_{\text{R}})$ and $\Gamma_{\text{slow}} = \text{min}(\Gamma_{\text{L}}, \Gamma_{\text{R}})$. For devices in the regime where $\Gamma\ll k_{\text{B}}T_0$ we use the lower bound $\Gamma_{\text{lower}}$ as described above. Since the maximum power factor in this regime is independent of the asymmetry between the fast and slow tunnel coupling (see Figure S21), the measured power factor for these devices are expected to fall on the black dashed line corresponding to $\chi = 1$. For the device where $\Gamma\ll k_{\text{B}}T_0$ G, B, E, F and C, we observe an increase of $(S^2 G)_\text{max}$ with increasing $\Gamma_{\text{lower}}$ due to the power factor being proportional to $G_{\text{max}}$. As $\Gamma$ approaches $k_{\text{B}}T_0$ the power factor reaches a maximum $ S^2G \approx \frac{1}{2.2} \times k_\text{B}^2/h$ for $\Gamma \approx 2.2 k_\text{B}T_0$. Devices D and Q were measured close to this maximum, as was the C$_{60}$ molecule measured by Kim \textit{et al.} at 100~K denoted `Kim' in Figure 4b.\cite{Kim2014} While devices D and `Kim' have a power factor close to the theoretical limit, for device Q $(S^2 G)_\text{max}$ is several orders of magnitude lower as a result of the asymmetric coupling $\chi \approx 10^4$ in this device. For $\Gamma\gg k_{\text{B}}T_0$ the maximum power factor is expected to decrease with increasing $\Gamma$ as the lifetime broadening reduces the Seebeck coefficient. No devices where measured in this regime.

Based on our finding, we conclude that there are three desiderata for achieving high thermoelectric performance in molecular nanodevices. First, the molecular energy levels need to align closely with the Fermi level of the electrodes since the Seebeck coefficient is maximum for $E$ close to the centre of the transmission resonance. Second, the tunnel coupling needs to be such that the lifetime of the transmission resonance is comparable to $k_{\text{B}}T_0$ at the operating temperature. Third, the tunnel couplings to the left and right electrode need to be equal to achieve a maximum power factor.

To summarise, we have fabricated thermoelectric nanodevices in which fullerene molecules are anchored between graphene source and drain leads. We demonstrate that by applying a thermal bias across the junction we can measure a gate dependent thermoelectricity. Our results show that by carefully tuning the transmission of a molecular junction towards sharp isolated resonance features, high power factors can be achieved approaching the theoretical limit of a thermally and lifetime broadened Coulomb peak. These results are relevant for the development of organic thermoelectric materials and our approach could also be applied to test hypotheses about the thermoelectric properties of molecules exhibiting quantum interference effects\cite{Finch2009} and spin caloritronics\cite{Wang2010}.

\section*{Methods}
\subsection*{Device fabrication}
Our devices are fabricated from single-layer CVD-grown graphene, which we transfer onto a Si/300 nm SiO$_2$ wafer with prepatterned 10 nm Cr/70 nm Au contacts and microheater. We pattern the graphene into a bow-tie shape (see Figure \ref{fig:Figure1}a) using standard electron beam lithography and O$_2$ plasma etching. The channel length $L$ of the devices and the width $W$ of the narrowest part of the constriction are 3.5~$\mu$m and 200~nm, respectively. To narrow down the constriction or form a nanogap we use a feedback-controlled electroburning technique in air\cite{Gehring2016} using an ADWin Gold II card with a 30~kHz sampling rate. Electroburning cycles are repeated until a critical resistance of 500~M$\Omega$ is reached.
\subsection*{Scanning thermal microscopy temperature measurements} 
	This method uses a temperature sensitive calibrated microfabricated probe with an apex of a few tens of nm that is brought in direct solid-solid contact with the sample. The SThM response $V_\text{t}$ is a linear function of the local sample temperature $T_\text{s}$. For a flat sample surface and constant tip-surface thermal resistance (that is the case when the tip is in contact with the same material – e.g. SiO$_2$) it allows to directly map a 2D distribution of the temperature increase in the vicinity of the micro-heater $\Delta T_\text{s}$, as well as to obtain an absolute value of the sample temperature increase due to micro-heater actuation using the following two quantitative methods: 1) In the “null-method” the probe apex temperature $T_\text{a}$ is varied, as the probe is brought repeatedly into contact with the sample. The value at which no change in the probe response $V_\text{t}$ occurs corresponds to $T_\text{s} = T_\text{a}$, which provides an absolute temperature measurement with an error of about 15 \% (see Chapter 2 and 3 Supporting Information for details).
	2) In the SThM “addition” method the sample is heated both by the micro-heater as well as by the calibrated raise in the temperature of the sample stage, allowing to perform measurements under vacuum and variable sample temperatures (see Chapter 2 and 3 Supporting Information for more details). These measurements show good correlation of the experimentally measured temperature maps with the finite-elements models. SThM measurements under ambient conditions were performed using a commercial SPM (Bruker MultiMode with Nanoscope E controller) and a custom-built SThM modified AC Wheatstone bridge. A resistive SThM probe (Kelvin Nanotechnology, KNT-SThM-01a, 0.3~N/m spring constant, $<100$~nm tip radius) served as one of the bridge resistors allowing precise monitoring of the probe AC electrical resistance at 91~kHz frequency via lock-in detection of the signal (SRS Instruments, SR830) as explained elsewhere\cite{Tovee2012}. Surface temperature maps were obtained at varying DC current to the probe that generated variable Joule heating of the probe tip. Several driving currents were used ranging from 0.10 to 0.40~mA leading to excess probe temperatures up to 34~K. The probe temperature - electrical resistance relation was determined employing a calibrated Peltier hot/cold plate (Torrey Pines Scientific, Echo Therm IC20) using a ratiometric approach (Agilent 34401A)\cite{Tovee2012}. The double-scan technique was used with different probe driving currents in order to obtain quantitative measurements of the surrounding and of the heater temperature\cite{Menges2012}. Laser illumination on the probe (on the order of 5~K) added to the Joule heating and was accounted via measurement of corresponding probe resistance change. SThM thermal mapping was performed with a set-force below 15~nN during imaging to protect the tip and the sample from damage.

\subsection*{Electric and thermoelectric transport measurements}
Graphene nano-structures were characterised in an Oxford Instruments Triton 200 dilution refrigerator with 20~mK base temperature. All measurements on C$_{60}$ junctions were performed in a liquid nitrogen dip-stick setup.
Electrical DC transport measurements were performed using low-noise DC electronics (Delft box). To measure the thermoelectric properties of nano-structures we used the $2f$ method\cite{Small2003}. To this end an AC heater voltage $V_\text{heat}(f)$ with frequency $f$ was applied to the micro-heater using a HP33120a arbitrary waveform generator. The thermovoltage was measured with a SR560 voltage pre-amplifier and a SRS830 lock-in amplifier at a frequency $2f$ (see Chapter 7 Supporting Information for more details).

\section*{Supporting Information Available}
Calibration of the heater: Resistance method; Calibration of the heater: Scanning thermal microscopy; Calibration of the heater: Vacuum and variable temperature Scanning thermal microscopy; Calibration of the heater: COMSOL simulations; Supporting thermoelectric data: C$_{60}$ junctions; Maximum power factor; Details on thermovoltage measurements; Error analysis.

\section*{Acknowledgements}
We thank the Royal Society for a University Research Fellowship for J.H.W. This work is supported by the UK EPSRC (grant nos. EP/K001507/1, EP/J014753/1, EP/H035818/1, EP/K030108/1, EP/J015067/1 and EP/N017188/1). O.K. acknowledges EU QUANTIHEAT FP7 no 604668 and EPSRC EP/G015570/1 for funding instrumentation in Lancaster. P.G. thanks Linacre College for a JRF. This project/publication was made possible through the support of a grant from Templeton World Charity Foundation. The opinions expressed in this publication are those of the author(s) and do not necessarily reflect the views of Templeton World Charity Foundation. The authors would like to thank J. Sowa, H. Sadeghi and C. Lambert for helpful discussions.

\section*{Competing financial interests}
The authors declare no competing financial interests.

\bibliography{scibib}

\bibliographystyle{achemso}

\end{document}